\documentstyle[aps,manuscript,psfig]{revtex}
\begin{document}
\draft
\begin{title}
{\bf Driven cavity flow: from molecular dynamics to  
continuum hydrodynamics} 
\end{title}
\author{Tiezheng Qian\footnote{To whom correspondence should be addressed. 
E-mail: maqian@ust.hk} and Xiao-Ping Wang}
\address{Department of Mathematics,
Hong Kong University of Science and Technology,\\
Clear Water Bay, Kowloon, Hong Kong, China} 
\maketitle

\begin{abstract}
Molecular dynamics (MD) simulations have been carried out to investigate 
the slip of fluid in the lid driven cavity flow where the no-slip 
boundary condition causes unphysical stress divergence.
The MD results not only show the existence of fluid slip
but also verify the validity of the Navier slip boundary condition.
To better understand the fluid slip in this problem,
a continuum hydrodynamic model has been formulated
based upon the MD verification of the Navier boundary condition 
and the Newtonian stress.
Our model has no adjustable parameter because all the material parameters 
(density, viscosity, and slip length)
are directly determined from MD simulations.
Steady-state velocity fields from continuum calculations are in quantitative  
agreement with those from MD simulations, from the molecular-scale
structure to the global flow.
The main discovery is as follows. 
In the immediate vicinity of the corners where moving and fixed 
solid surfaces intersect, there is a core partial-slip region where
the slippage is large at the moving solid surface
and decays away from the intersection quickly. 
In particular, the structure of this core region is nearly independent 
of the system size.
On the other hand, for sufficiently large system,
an additional partial-slip region appears where the slippage 
varies as $1/r$ with $r$ denoting the distance from the corner 
along the moving solid surface.
The existence of this wide power-law region is in accordance with 
the asymptotic $1/r$ variation of stress and the Navier boundary condition.   
\end{abstract}
\narrowtext

\section{Introduction}

A crucial ingredient in the continuum hydrodynamics is the
boundary condition of fluid flow past a solid surface.
The no-slip boundary condition,
i.e., zero relative velocity between the fluid and solid at the
interface, is a core concept in fluid mechanics \cite{batchelor}.
In molecular dynamics (MD) simulations, however, 
a small amount of relative slip between the fluid and the solid surface 
is generally detected \cite{Th-Rob,nbc,barrat,ckb}. 
Such slip can be accounted for by the Navier boundary condition (NBC),
whereby the slip velocity is proportional to the tangential viscous 
stress and the degree of slip is measured by a slip length 
\cite{Th-Rob,nbc,barrat,ckb}. 
As the relative slip is extremely small in macroscopic flows, the NBC 
is practically indistinguishable from the no-slip boundary condition 
in most situations. 

When applied to the immiscible two-phase flow of moving contact line, 
where a fluid-fluid interface intersects the solid wall,
the no-slip boundary condition would cause non-integrable diverging stress
and unphysical infinite dissipation, which directly imply the breakdown of 
the no-slip boundary condition \cite{dussan}. In the past two decades, 
MD simulations have shown fluid slip in the molecular-scale vicinity 
of the moving contact line \cite{koplik,robbins}. 
Recent evidences have shown the slip velocity profile obtained from MD
simulations to be accountable by the generalized Navier boundary
condition \cite{qws}, in which the slip velocity is
proportional to the total tangential stress --- the sum of the
viscous stress and the uncompensated Young stress; the latter
arises from the deviation of the fluid-fluid interface from its
static configuration.

The no-slip boundary condition also runs into trouble 
when applied to the driven cavity flow, where a rigid plane 
slides steadily over another, with a constant inclination angle 
\cite{batchelor,corner-flow}. 
This geometry readily shows that the no-slip boundary condition
would cause the same non-integrable diverging stress as in the problem 
of moving contact line. Near the corner where fixed and moving
solid surfaces intersect, the velocity variation becomes very fast 
because two different velocities are assumed at the two solid surfaces. 
Moreover, a velocity discontinuity occurs at the corner if the no-slip 
boundary condition is everywhere applied.
This is the origin of the non-integrable $1/r$ stress at $r\rightarrow 0$. 
A slip boundary condition is therefore imperative.

Koplik and Banavar first used MD simulations to explore the small-scale
structure of the driven cavity flow \cite{corner-flow}. Their results
indicate that slip occurs in the corner region. To uncover the slip mechanism, 
they measured the microscopic tangential stress at the fluid-solid interface. 
This stress measurement led to the conclusions that local non-Newtonian region 
exists in a low-shear and otherwise Newtonian flow (at low Reynolds number) 
and that the NBC is not valid.

The purpose of this paper is to uncover the slip boundary condition and
formulate a continuum hydrodynamic model for the driven cavity flow, 
from which we will answer the intriguing question: 
In a mesoscopic or macroscopic system, 
what is the slip profile which consistently interpolates
between the inevitable slippage in the immediate 
vicinity of the corner and the no-slip boundary condition 
that must hold at mesoscopic/macroscopic regions far away?
Success would lead to new understanding to slip and dissipation in 
restricted geometries \cite{restricted}.

The continuum hydrodynamic modeling requires a slip boundary condition 
and a momentum transport equation. We have carried out MD simulations  
similar to those performed for the moving contact line problem \cite{qws}.
In contrast to the conclusions in Ref. \cite{corner-flow},
the NBC is found to be governing the slip of fluid 
relative to the solid. The shear stress is verified 
to be Newtonian in the vicinity of the corner where velocity
and stress variations are extremely large. Technically,
we denote a molecular boundary layer of fluid at the fluid-solid interface.
We then perform stress measurement using a method that is reliable
near the interface \cite{qws}. Velocity and stress data
collected at the boundary layer provide molecular evidence for the validity 
of the NBC. We emphasize that unlike the Couette flow simulated in 
Ref. \cite{nbc,barrat,ckb}, here the driven cavity flow requires 
a local verification of the NBC, because the slip velocity varies along 
the solid surface and this variation becomes very fast in the corner region.
As for the Newtonian behavior, local velocity and shear stress 
are measured everywhere in the fluid, showing that the shear stress 
is proportional to the local shear rate. 

Based upon the conclusions drawn directly from the simulated cavity flow, 
a continuum hydrodynamic model has been formulated,
comprising the Navier-Stokes equation and the NBC. 
With all the material parameters directly determined from MD simulations, 
our model has {\it no} adjustable parameter. 
Numerical calculations have been carried out to produce continuum results 
for comparison with MD results. 
It is shown that in a wide range of Reynolds number, 
MD and continuum flow fields agree well, 
from the molecular-scale corner region (with large slip) 
to the large-scale outer region (with vanishing slip in a large system).
The largest Reynolds number ever reached is $\approx 50$, 
at which deviation from the Stokes flow is clearly noted
(see Sec. \ref{large-reynolds}).

The paper is organized as follows. We first describe the details of the MD 
simulations in Sec. II. We then outline in Sec. III our MD approach 
to the verification of the Navier slip boundary condition. 
The MD results, later used for hydrodynamic modeling, are presented 
in Sec. IV. A continuum hydrodynamic model is formulated in Sec. V. 
The numerical algorithm is also briefly described.
In Sec. VI there is a systematic comparison of the MD and continuum 
hydrodynamics results. The paper is concluded in Sec. VII with a few remarks.

\section{Molecular dynamics simulations}

The purpose of carrying out MD simulations is threefold:
(1) To uncover the boundary condition governing the driven cavity flow
(Secs. \ref{MD-slip} and \ref{MD-results});
(2) To determine the material parameters (e.g., viscosity and slip length)
in our hydrodynamic model (Sec. \ref{continuum-model});
(3) To produce flow fields for comparison with the continuum hydrodynamic 
solutions (Sec. \ref{MD-continuum-comparison}).

\begin{figure}
\centerline{\psfig{figure=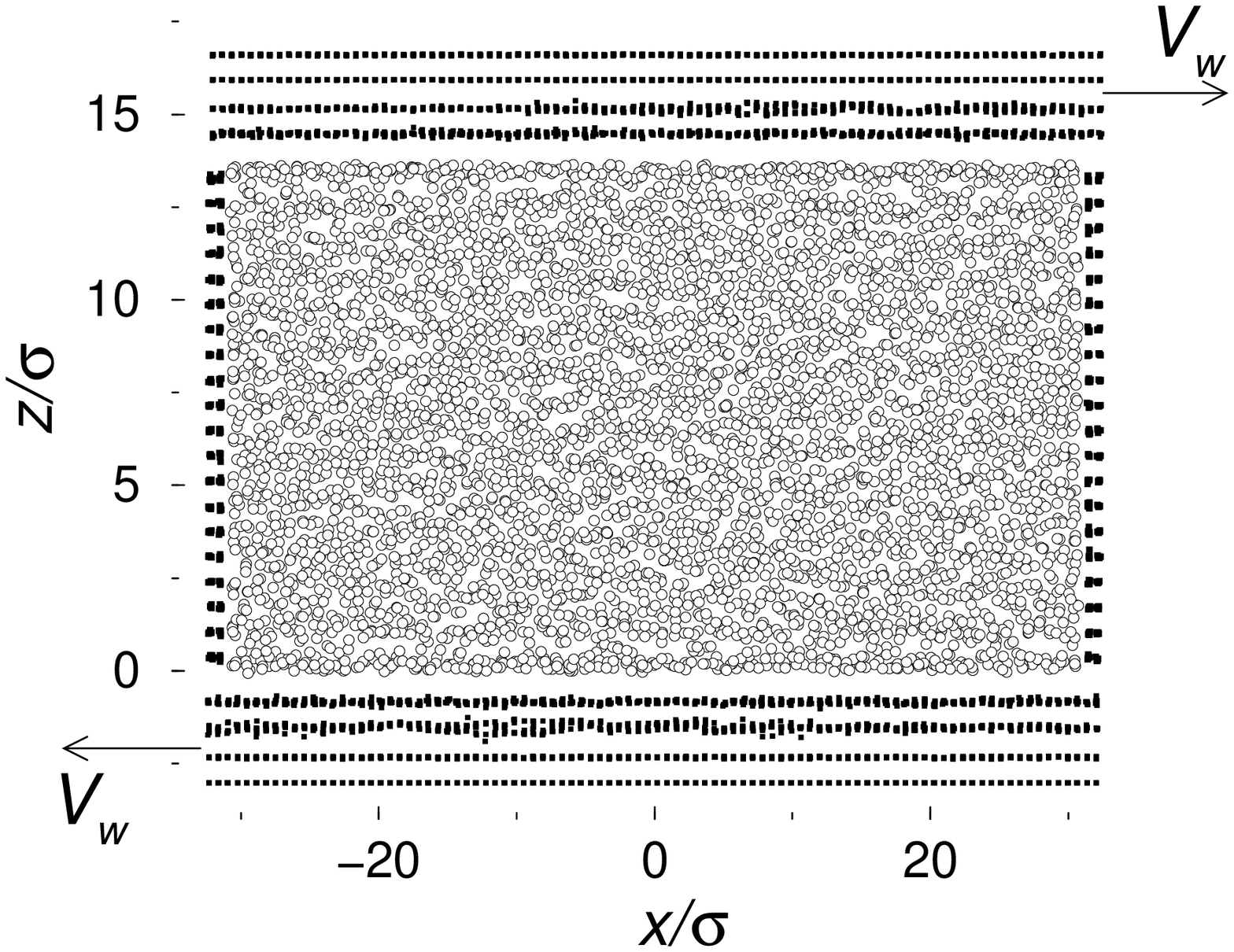,height=12cm}}
\bigskip
\caption{Geometry of MD simulations. The empty circles 
indicate the instantaneous molecular positions of the 
fluid projected onto the $xz$ plane. The solid squares
denote the wall molecules. 
}\label{fig-geo}
\end{figure}

We consider a single fluid confined in a two-dimensional (2D) cavity formed 
by two horizontal walls in the $xy$ plane and two vertical walls 
in the $yz$ plane (see Fig. \ref{fig-geo}) \cite{corner-flow}. 
The cavity measures $L$ along $x$ and 
$H$ along $z$, with the periodic boundary condition applied along $y$.
The fluid is sheared by moving the upper and lower walls 
with the same speed $V_w$ along the $\pm x$ directions, respectively.
Each of the four walls is constructed by two to four [001] planes of 
an fcc lattice, with each wall molecule attached to the lattice site 
by a harmonic spring. The mean-squared displacement of wall molecules
is controlled to obey the Lindemann criterion.
Interaction between the fluid molecules separated by a distance $r$
is modeled by a Lennard-Jones (LJ) potential
$$U_{ff}=
4\epsilon\left[\left(\displaystyle\frac{\sigma}{r}\right)^{12}
-\left(\displaystyle\frac{\sigma}{r}\right)^6\right],$$ 
where $\epsilon$ and $\sigma$ 
are the energy scale and range of interaction, respectively.
The wall-fluid interaction is modeled by a modified LJ potential
$$U_{wf}=
4\epsilon_{wf}\left[\left(\displaystyle\frac{\sigma_{wf}}{r}\right)^{12}
-\delta_{wf}\left(\displaystyle\frac{\sigma_{wf}}{r}\right)^6\right],$$
with energy and range parameters $\epsilon_{wf}$ and $\sigma_{wf}$, 
and a $\delta_{wf}$ for tuning the wetting property of the fluid.
Both $U_{ff}$ and $U_{wf}$ are truncated at $2.5\sigma$.
In our simulations, the density of fluid $\rho$ equals to $0.81\sigma^{-3}$,
the density of wall $\rho_w$ equals to $1.86\sigma^{-3}$ (which determines 
the wall lattice constant), the mass of wall molecule $m_w$  
equals to the mass of fluid molecule $m$,
the parameters in $U_{wf}$ are $\epsilon_{wf}=1.16\epsilon$, 
$\sigma_{wf}=1.04\sigma$, and $\delta_{wf}=1.0$, and the temperature 
$T$ is fixed at $2.8\epsilon/k_B$. 
The values of $V_w$, $L$ and $H$ are varied as external conditions
in different simulations. 
The steady-state flow fields are obtained from time averages over
$10^4$ to $10^6\tau$ where $\tau$ is the atomic time scale
$\sqrt{m\sigma^2/\epsilon}$.
We have also performed similar simulations for other temperatures 
ranging from $1.2\epsilon/k_B$ to $3.0\epsilon/k_B$.  
The MD velocity profiles can always be reproduced by our continuum model,
with material parameters directly determined from MD simulations.
This is due to the fact that the fluid remains to be Newtonian 
and the slip length is a constant at a given temperature.
Details will be presented in Sec. \ref{temperature-effect}.

We denote the region within $z_0=0.425\sigma$ of the fluid surface
the boundary layer (BL). 
It must be thin enough to ensure sufficient precision for 
measuring the slip velocity at the solid surface,
but also thick enough to fully account for the tangential 
wall-fluid interaction force.
The wall force can be singled out by separating the force on 
each fluid molecule into wall-fluid and fluid-fluid components. 
The fluid molecules in the BL, being close to the solid wall, 
can detect the discrete structure of the wall (the `roughness'
of the wall potential \cite{nbc}). When coupled with kinetic collisions 
with the wall molecules, there arises a nonzero tangential 
wall force density $g_x^w$ that is sharply peaked at 
$z\approx z_0/2$ and vanishes beyond $z\approx z_0$. Here 
the subscript $x$ in $g_x^w$ and the $z$ coordinates are for the BL 
at the lower fluid-solid interface (same below), 
with the understanding that the same physics holds at 
the other three fluid-solid interfaces. 
From the force density $g_x^w$, we define the tangential wall force 
per unit area as $G_x^w(x)=\int_0^{z_0}dz {g}_x^w(x,z)$, 
which is the total tangential wall force accumulated across the BL.

Spatial resolution along the $x$ and $z$ directions is achieved
by evenly dividing the sampling region into bins, each
$\Delta x=0.85\sigma$ by $\Delta z =0.425\sigma$ in size. 
The slip velocity $v_x^{slip}$ in the lower/upper BL is obtained as 
the time average of fluid molecules' velocities in each BL bin, 
measured with respect to the moving wall ($v_x^{slip}=v_x+V_w$
in the lower BL, or $v_x^{slip}=v_x-V_w$ in the upper BL); 
the tangential wall force $G_x^w$ in the lower/upper BL is obtained 
from the time average of the total tangential wall force experienced 
by the fluid molecules in each BL bin, divided by the bin area 
in the $xy$ plane; 
the fluid stress component $\sigma_{xx(zx)}$ is obtained from 
the time averages of the kinetic momentum transfer plus the fluid-fluid 
interaction forces across the constant-$x(z)$ bin surfaces, 
and the fluid velocity component $v_{x(z)}$ is measured as 
the time-average of that component within each bin.
In particular, we have directly measured the fluid-fluid interaction 
forces across bin surfaces to obtain the contribution of intermolecular 
forces to the fluid stress, because the validity of 
the Irving-Kirkwood stress expression was noted to be not justified 
at a fluid-fluid or fluid-solid interface \cite{irving-kirkwood}.
The technical details for the stress measurement near a fluid-fluid or 
fluid-solid interface may be found in Appendix B of Ref. \cite{qws}.

\section{Slip boundary condition}\label{MD-slip}

Figure \ref{fig-slip} shows the MD evidence for the existence of slip.  
It is seen that the slippage along $x$ becomes quite large near the corner,
as already observed in Ref. \cite{corner-flow}. 
In particular, the fluid undergoes near-complete slip 
in approaching the corner, regardless of the system size. 
(By near-complete, we mean $|v_x^{slip}|$ approaches $V_w$.) 
Far away from the corner 
(for fluid that extends long enough along $x$, i.e., $L\gg H$), 
the flow is not perturbed by the vertical walls due to viscous damping, i.e.,
the uniform shear flow prevails and a small constant slip is detected.

The NBC is the simplest alternative of the no-slip boundary condition. 
It states that the amount of slip is proportional to the tangential 
fluid stress at the solid surface. For a Newtonian fluid, 
the tangential viscous stress is proportional to the shear rate. 
Consequently, the NBC becomes that the amount of slip is proportional to 
the shear rate $\dot{\gamma}$, i.e., $v_x^{slip}=l_s\dot{\gamma}$, where
the proportionality constant $l_s$ is the slip length \cite{nbc,barrat,ckb}. 
Physically, a nonzero slip length arises from 
the unequal wall and fluid densities, the weak wall-fluid interaction, 
and the high temperature. 
Together, they prevent the epitaxial locking of fluid layer(s) to 
the solid wall, and thus allow slip to occur. 
A recent study shows that fluid flow in carbon nanopores is characterized 
by a large slip length \cite{sokhan}.

The verification of the NBC in the driven cavity flow, where slip velocity
varies along the solid wall, consists of three stages.
(i) We show that $v_x^{slip}$ is proportional to $G_x^w$,
the local tangential wall force per unit area.
(ii) We show that $G_x^w$ is balanced by the tangential fluid force 
$G_x^f$. This force balance is necessary because 
inertial effects are negligible in the BL of molecular thickness.
Accordingly, $v_x^{slip}$ is also proportional to $G_x^f$.
Here we emphasize that our stress measurement scheme has been designed 
to obtain the tangential fluid force correctly.
(iii) We show that the shear stress is Newtonian. 
From (ii) and (iii) the slip length $l_s$ can be defined and measured.

\section{MD results for hydrodynamic modeling}\label{MD-results}

As shown in Fig. \ref{fig-wall}, 
the tangential wall force per unit area $G_x^w$ 
is proportional to the local slip velocity $v_x^{slip}$:
\begin{equation}\label{gwslip}
G_x^w(x)=-\beta v_x^{slip}(x),
\end{equation}
where the proportionality constant $\beta$ is the slip coefficient.
To see if nonlinearity would arise for large $V_w$, MD simulations 
have been performed at $V_w\sim 1.0\sqrt{\epsilon /m}$. 
Figure \ref{fig-wall} shows that for the wall-fluid interaction used here,  
nonlinearity in the slip boundary condition is accessible when
$|v_x^{slip}|>0.5\sqrt{\epsilon /m}$. According to the nonlinear effect 
discovered by Thompson and Troian \cite{nbc}, 
the slip length $l_s=\eta/\beta$ increases with the increasing slip velocity 
when the latter is sufficiently large. This is seen in
Fig. \ref{fig-wall} where $\beta$ decreases with the increasing $|v_x^{slip}|$.

\begin{figure}
\centerline{\psfig{figure=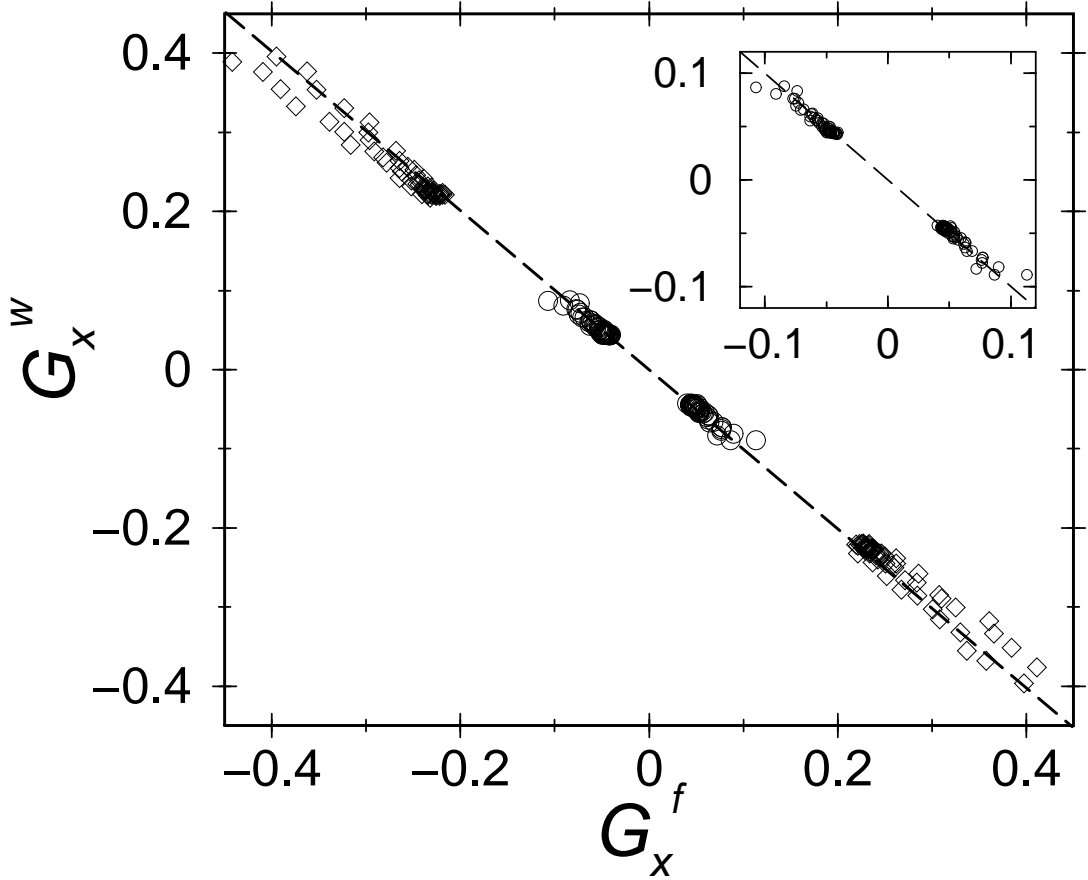,height=12.0cm}}
\bigskip
\caption{The tangential wall force $G_x^w$
(in the unit of $\epsilon/\sigma^3$) 
is plotted as a function of the tangential fluid force $G_x^f$
(in the unit of $\epsilon/\sigma^3$)
for the lower ($G_x^f>0$) and upper ($G_x^f<0$) BL's.
The two cases shown here are of the same $L=61.3\sigma$ and
$H=13.6\sigma$ but different $V_w$.
The circles and diamonds denote the cases of 
$V_w=0.25\sqrt{\epsilon /m}$
and $1.25\sqrt{\epsilon /m}$, respectively. 
The dashed line has the slope of $-1$, indicating
$G_x^w+G_x^f=0$. 
The inset shows an enlarged plot for 
$V_w=0.25\sqrt{\epsilon /m}$.
}\label{fig-balance}
\end{figure}

The BL is thin enough to make the inertial term negligible
in momentum equation ($m\rho V_w z_0/\eta\ll 1$). 
It follows that the tangential wall force $G_x^w$ is balanced by 
the tangential fluid force per unit wall area, $G_x^f$, 
i.e., $G_x^w(x)+G_x^f(x)=0$, as shown in Fig. \ref{fig-balance}.
Here $G_x^f$ is of the form
\begin{equation}\label{gf-expression}
G_x^f(x)=\sigma_{zx}(x,z_0)+\partial_x\int_0^{z_0}dz\sigma_{xx}(x,z),
\end{equation}
coming from $$G_x^f(x)=\int_0^{z_0}dz
[\partial_x\sigma_{xx}(x,z)+\partial_z\sigma_{zx}(x,z)],$$
together with the fact that $\sigma_{zx}(x,0)=0$.
(More strictly, $\sigma_{zx}(x,0^-)=0$ because there is no fluid
below $z=0$, hence no momentum transport across $z=0$.)
It follows from Eq. (\ref{gwslip}) and the tangential force balance that
\begin{equation}\label{gfslip}
G_x^f(x)=\beta v_x^{slip}(x).
\end{equation}

It is worth emphasizing that the normal stress $\sigma_{xx}$ 
in the BL exhibits extremely large variation in the partial-slip region  
close to the corner, where $v_x^{slip}$ varies quickly along $x$. 
In fact, the normal stress variation is so large that 
a small fluid density difference is even noted 
between the low- and high-pressure regions. Quantitatively, 
the BL-integrated normal stress $\int_0^{z_0}dz\sigma_{xx}(x,z)$
is essential to reaching Eq. (\ref{gfslip}), 
because close to the corner, the contribution of 
$\partial_x\int_0^{z_0}dz\sigma_{xx}(x,z)$ to $G_x^f$ is 
of the same order as that of $\sigma_{zx}(x,z_0)$.
(In the region of uniform shear flow far away from the corner,
$\partial_x\int_0^{z_0}dz\sigma_{xx}(z)=0$.)

To summarize, in obtaining the Navier slip condition
Eq. (\ref{gfslip}), we first identify the BL and measure the tangential  
wall force therein to obtain Eq. (\ref{gwslip}).  We then measure the normal 
and tangential stresses $\sigma_{xx}$ and $\sigma_{zx}$ according to the
original definition of stress (because the Irving-Kirkwood expression is 
not reliable near the fluid-solid interface). We finally
calculate the tangential fluid force according to Eq. (\ref{gf-expression})
to verify the equation of BL force balance.

Equation (\ref{gf-expression}) is due to the finite thickness of 
the BL, in which a tangential wall force is sharply distributed 
along the solid surface normal. 
Nevertheless, it is a reasonable expectation that the fluid 
would experience almost the identical physical effect(s) 
from a wall force density ${G}_x^w\delta(z)$, concentrated strictly 
at $z=0$ with the same total wall force per unit area. 
Replacing a diffuse BL by a sharp BL can
considerably simplify the form of the boundary condition, 
because local force balance along $x$ then requires
$\partial_x\sigma_{xx}+\partial_z\sigma_{zx}=0$ away from $z=0$. 
Integration of this relation from $0^+$ to $z_0$ yields
$$\partial_x\int_0^{z_0}dz\sigma_{xx}(x,z)
+\sigma_{zx}(x,z_0)-{\sigma}_{zx}(x,0^+)=0.$$ 
A comparison with Eq. (\ref{gf-expression}) then relates $G_x^f$ to 
$\sigma_{zx}$ at surface: $G_x^f(x)=\sigma_{zx}(x,0^+)$. Therefore, 
$\sigma_{zx}$ changes from $\sigma_{zx}(x,0^-)=0$
to $\sigma_{zx}(x,0^+)=G_x^f(x)$ at $z=0$, leading to
$$(\nabla\cdot{\mbox{\boldmath$\sigma$}})\cdot\hat{\bf x}
=G_x^f\delta(z).$$ 
This tangential fluid force density is in balance with
the tangential wall force density ${G}_x^w\delta(z)$.
Now the BL is from $0^-$ to $0^+$,
instead of from $0$ to $z_0$ as in the diffuse case. Correspondingly, 
the NBC becomes
\begin{equation}\label{sharp-boundary-nbc}
\sigma_{zx}(x,0)=\beta v_x^{slip}(x)
\end{equation}
in the sharp boundary limit.

\begin{figure}
\centerline{\psfig{figure=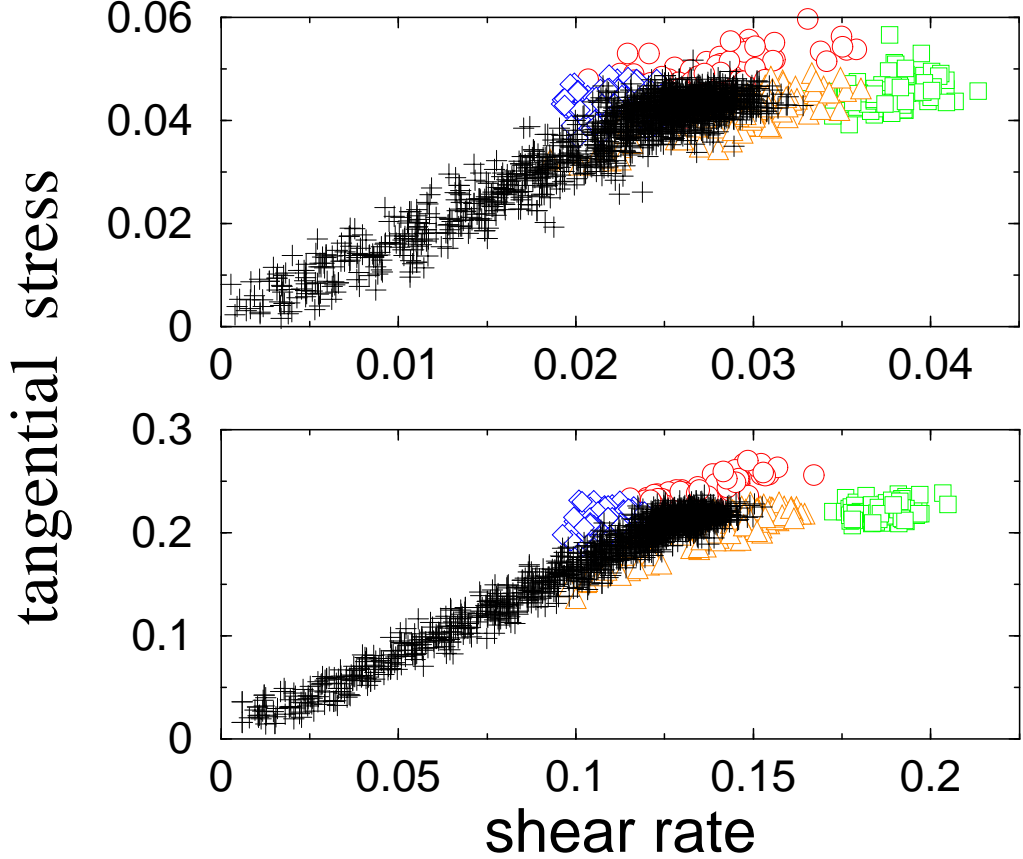,height=12cm}}
\bigskip
\caption{Newtonian response of shear viscous stress.
The tangential stress $\sigma_{zx}$ 
(in the unit of $\epsilon/\sigma^3$) is plotted as
a function of the shear rate $\partial_z v_x+\partial_x v_z$
(in the unit of $\tau^{-1}$).
The upper panel shows the case of $L=61.3\sigma$,
$H=13.6\sigma$, and $V_w=0.25\sqrt{\epsilon /m}$;
the lower panel shows the case of $L=61.3\sigma$,
$H=13.6\sigma$, and $V_w=1.25\sqrt{\epsilon /m}$.
In each panel, the circles denote the data collected at
the levels of $z=z_0$ and $z=H-z_0$, 
the squares denote the data collected at
the levels of $z=2z_0$ and $z=H-2z_0$, 
the diamonds denote the data collected at
the levels of $z=3z_0$ and $z=H-3z_0$, 
the triangles denote the data collected at
the levels of $z=4z_0$ and $z=H-4z_0$, and the
pluses denote the data collected at all other levels.
At each $z$ level, $\sigma_{zx}$ shows a linear dependence on 
$\partial_z v_x+\partial_x v_z$. The ratio of 
$\sigma_{zx}$ to $\partial_z v_x+\partial_x v_z$
varies across the first four $z$ levels away from the wall,
and becomes a $z$-independent constant at levels deeper 
in the fluid. Note that the ratio of 
$\sigma_{zx}$ to $\partial_z v_x+\partial_x v_z$
remains to be unchanged from $V_w=0.25\sqrt{\epsilon /m}$
to $V_w=1.25\sqrt{\epsilon /m}$. (The scales for 
$\sigma_{zx}$ and $\partial_z v_x+\partial_x v_z$ 
in the lower panel are five times larger than those
in the upper panel, corresponding to the five times 
difference between the two values of $V_w$.)
}\label{fig-newton}
\end{figure}

It remains to be seen if the fluid is Newtonian, that is, if 
the viscous stress tensor is proportional to the rate of strain tensor.
In particular, we need to find out if the tangential stress $\sigma_{zx}$ 
is still proportional to the local shear rate $\dot{\gamma}$ as 
the fluid-solid interface is approached.
We have measured both $\sigma_{zx}$ and $\dot{\gamma}$ at 
$z=z_0,2z_0,\cdot\cdot\cdot$
($z=z_0$ is the top surface of the lower BL).
As shown in Fig. \ref{fig-newton}, the ratio of $\sigma_{zx}$ to 
$\dot{\gamma}$ is indeed a constant at each $z$ level, 
but this constant varies a little along $z$ near the wall, due to the 
short-range density modulation induced by the rigid wall \cite{israelachivili}. 
Far away from the wall, the fluid density
approaches a $z$-independent constant, and so does the ratio of 
$\sigma_{zx}$ to $\dot{\gamma}$.
To see if non-Newtonian response would arise for large $V_w$, 
MD simulations have been performed at $V_w\sim 1.0\sqrt{\epsilon/m}$. 
Figure \ref{fig-newton} shows that the viscosity remains 
to be a constant for $V_w$ as large as $1.25\sqrt{\epsilon/m}$.
This is due to the fact that for $V_w\sim 1.0\sqrt{\epsilon/m}$
and $H>10\sigma$, the shear rate is $\sim 0.1\tau^{-1}$ or smaller,
whereas non-Newtonian response of bulk fluid is expected for
$\dot{\gamma}\ge 2\tau^{-1}$.

From Eq. (\ref{gfslip}) and the sharp boundary limit
$G_x^f(x)=\sigma_{zx}(x,0)$, we have Eq. (\ref{sharp-boundary-nbc}).
As the shear stress is verified to be Newtonian, i.e.,
$\sigma_{zx}=\eta\dot{\gamma}$, we obtain
the commonly used hydrodynamic NBC
\begin{equation}\label{nbc}
v_x^{slip}(x)=l_s\partial_zv_x(x,0),
\end{equation}
with the slip length $l_s=\eta/\beta$.
According to the sharp boundary limit involved in obtaining 
Eq. (\ref{nbc}), we should compare the MD tangential fluid force $G_x^f(x)$ 
with the continuum tangential viscous stress $\eta\partial_zv_x(x,0)$ 
at surface.
Practically, a comparison between the MD and continuum 
profiles of $v_x^{slip}$ would suffice because
$G_x^f=\beta v_x^{slip}$ in the MD and 
$\eta\partial_zv_x(0)=\beta v_x^{slip}$ in the continuum hydrodynamics.

\section{Continuum hydrodynamic model}\label{continuum-model}
A continuum hydrodynamic model has been formulated for the driven  
cavity flow from our knowledge of the NBC and the Newtonian stress.
An explicit scheme has been designed to solve the hydrodynamic model, 
comprising the Navier-Stokes equation and the NBC.
Material parameters include the fluid density $\rho$ ($=0.81\sigma^{-3}$), 
the viscosity $\eta$ ($=1.65\sqrt{\epsilon m}/\sigma^2$), 
and the slip length $l_s=\eta/\beta$ ($=2.3\sigma$), all directly 
determined from MD simulations.
Thus our model has {\it no} adjustable parameter.
Numerical calculations show that steady-state flow fields from
MD simulations can be quantitatively reproduced.

The 2D flow is governed by the Navier-Stokes equation
$$
m\rho\left[\displaystyle\frac{\partial{\bf v}}{\partial t}+
\left({\bf v}\cdot\nabla\right){\bf v}\right]=
-\nabla p+\eta\nabla^2{\bf v},
$$ 
with the incompressibility conditions 
$\nabla\cdot{\bf v}=0$, and the boundary conditions:
$v_z=0$ and $l_s^{-1} v^{slip}_x=-\partial_nv_x$ 
at the moving horizontal walls;
$v_x=0$ and $l_s^{-1} v^{slip}_z=-\partial_nv_z$ 
at the vertical walls ($n$ denotes the outward surface normal).

Our continuum model has six parameters, including 
the system dimensions $L$ along $x$ and $H$ along $z$, 
the speed of the moving horizontal walls $V_w$, 
the fluid density $\rho$, the viscosity $\eta$,
and the slip length $l_s$. 
Taking $H$ as the length unit, $V_w$ as the velocity unit,
and $\eta V_w/H$ as the pressure/stress unit, we are left with three    
dimensionless controlling parameters: 
the aspect ratio $L/H$, 
the dimensionless slip length ${l_s}/{H}$,
and the Reynolds number ${\cal R}={\rho V_wH}/{\eta}$. 
In the regime of small Reynolds number (Stokes flow) 
the only controlling parameters are $L/H$ and $l_s/H$. 

The finite-difference scheme used for solving the Navier-Stokes equation 
is a modified version of the Pressure-Poisson formulation 
given in Ref. \cite{liu,qws}, where the incompressibility condition 
is replaced by the pressure Poisson equation and a divergence-free
boundary condition for the velocity. Variable grids 
with better resolution near the corners are used to save computational cost.

\section{Comparison of MD and continuum results}\label{MD-continuum-comparison}

\subsection{Stokes flow}
For ${\cal R}\sim 1$ or smaller, nonlinear effects associated with
the inertial term are negligible. As a result, the dimensionless
steady-state solution for the flow field,  
${\bf v}({\bf r}/H)/V_w$, depends on $L/H$ and $l_s/H$ only. 
Figure \ref{fig-slip} shows the MD profiles of $v_x/V_w$
in the BL (the slip profiles), obtained from three simulations 
using the same system size ($L$ and $H$) 
but different wall speed $V_w$. It is seen that
the three MD profiles approach the same limiting profile 
\cite{note-nonlinear}. 
This is due to the fact that the Reynolds number ${\cal R}$ ranges from 
$0.067$ to $8.3$ in the three cases, thus justifies the Stokes-flow limit.
In Fig. \ref{fig-stokes} the MD profiles of $v_x/V_w$
at different $z$ levels, obtained from two of the three simulations
shown in Fig. \ref{fig-slip}, also indicate the Stokes-flow limit.

For comparison with the above MD results, the corresponding
continuum results are also plotted in Figs. \ref{fig-slip}
and \ref{fig-stokes}. They were calculated using the same set of 
material parameters $\rho$, $\eta$, and $l_s$ (see Sec. \ref{continuum-model}) 
under respective conditions for $L$, $H$, and $V_w$.
The continuum calculations involve no adjustable parameter, 
and the overall agreement is satisfactory.
A small discrepancy is noticed for the BL tangential velocity (or
slip velocity) in a small region close to the corner where the slip amount 
is relatively large and displays sharp decay. 
This is presumably due to the short-range density modulation induced by 
the rigid wall \cite{israelachivili}, given the short distance 
$H=13.6\sigma$ here \cite{subcontinuum}.
In fact, this discrepancy tends to be less noticeable for larger $H$.

\subsection{Temperature effects}\label{temperature-effect}
\begin{figure}
\centerline{\psfig{figure=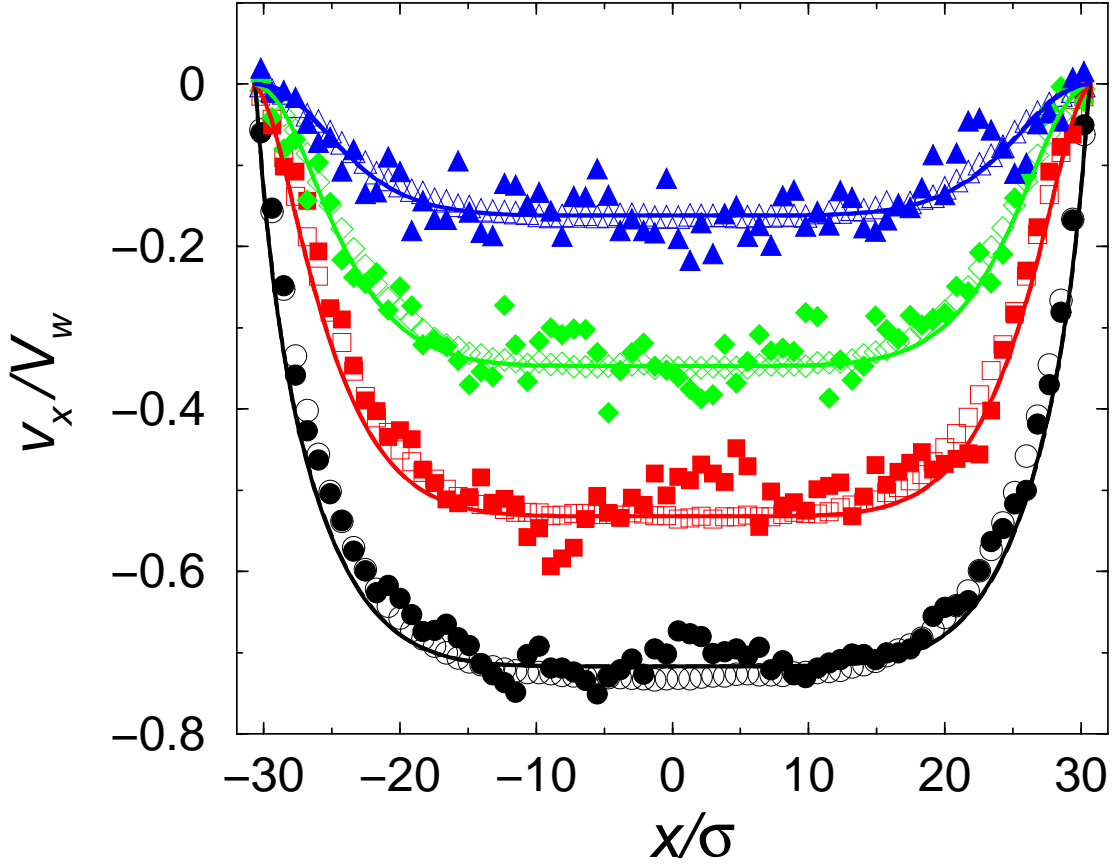,height=12.cm}}
\bigskip
\caption{Comparison between the MD and continuum results for two Stokes flows.
The symbols denote the MD profiles of $v_x/V_w$ at different $z$ levels,
obtained for the same $L=61.3\sigma$ and $H=13.6\sigma$ but different $V_w$. 
The empty and solid symbols represent the cases of $V_w=0.25\sqrt{\epsilon /m}$ 
and $0.01\sqrt{\epsilon /m}$, respectively.
The $v_x/V_w$ profiles are symmetric about the center plane $z=H/2$,
hence only the lower half is shown for $z=0.2125\sigma$ (circles), 
$1.9125\sigma$ (squares), $3.6125\sigma$ (diamonds), 
and $5.3125\sigma$ (triangles).
Large fluctuation is seen from the solid symbols for 
$V_w=0.01\sqrt{\epsilon /m}$ 
(smaller than the thermal velocity by two orders of magnitude),
although we averaged over $\sim 10^6\tau$ to reduce noise. 
The solid and dashed lines denote the corresponding continuum profiles for 
$V_w=0.25\sqrt{\epsilon /m}$ (${\cal R}=1.67$) and $0.01\sqrt{\epsilon /m}$ 
(${\cal R}=0.067$), respectively.
}\label{fig-stokes}
\end{figure}

\begin{figure}
\centerline{\psfig{figure=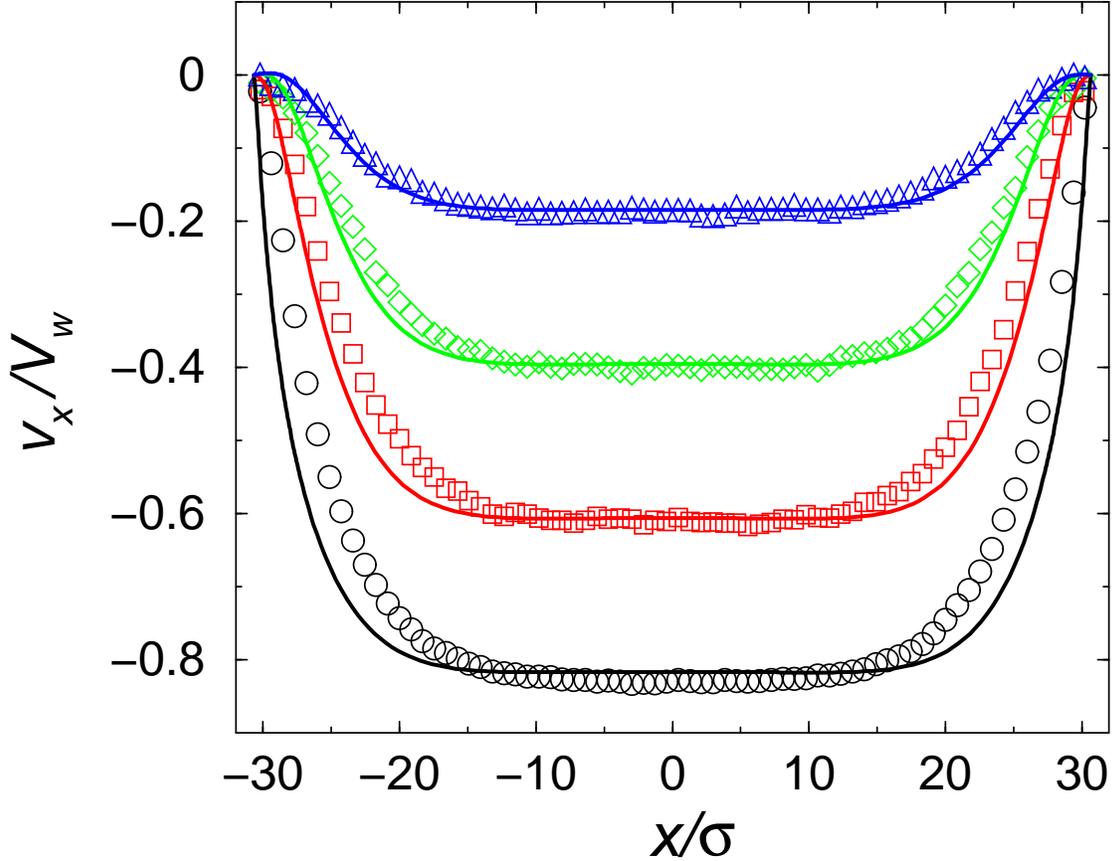,height=12.cm}}
\bigskip
\caption{Comparison between the MD and continuum 
results for a Stokes flow at temperature $T=1.4\epsilon/k_B$.
The symbols denote the MD profiles of $v_x/V_w$ at different $z$ levels,
obtained for $V_w=0.25\sqrt{\epsilon /m}$, $L=61.3\sigma$, and $H=13.6\sigma$.
The $v_x/V_w$ profiles are symmetric about the center plane $z=H/2$,
hence only the lower half is shown for $z=0.2125\sigma$ (circles), 
$1.9125\sigma$ (squares), $3.6125\sigma$ (diamonds), 
and $5.3125\sigma$ (triangles).
The solid lines denote the corresponding continuum profiles.
}\label{fig-temp}
\end{figure}

MD simulations have been carried out as well for temperatures other than 
$T=2.8\epsilon/k_B$. We find that for $T$ ranging from $1.2\epsilon/k_B$ 
to $3.0\epsilon/k_B$, the MD velocity profiles can always be reproduced 
by our continuum model, with material parameters directly determined from 
MD simulations. In Fig. \ref{fig-temp}, we show the MD profiles 
of $v_x/V_w$ at different $z$ levels, obtained from a simulation at
$T=1.4\epsilon/k_B$. The corresponding continuum results are also shown
for comparison. They were calculated using the material parameters 
$\rho=0.81\sigma^{-3}$, $\eta=1.7\sqrt{\epsilon m}/\sigma^2$, 
and $l_s=1.27\sigma$. (MD measurements show that the viscosity weakly
depends on the temperature, whereas the slip length is strongly 
temperature-dependent.)
It is noticed that in the partial-slip region close to the corner, 
the discrepancy here is a bit larger than that seen for $T=2.8\epsilon/k_B$
in Fig. \ref{fig-stokes}.
Again, this is caused by the short-range density modulation 
induced by the rigid wall \cite{israelachivili}, given the same 
short distance $H=13.6\sigma$ here. In particular, such near-surface
density modulation becomes more prominent as the temperature is lowered,
and that's why the agreement here is less satisfactory.

The uniform shear flow in the central region of the cavity is seen in 
Figs. \ref{fig-stokes} and \ref{fig-temp}. This part of the cavity flow 
is simply described by the Navier-Stokes equation $\partial_z^2v_x=0$ 
plus the NBC $v_x^{slip}=\pm l_s\partial_z v_x$ at $z=0$ and $H$, 
from which a constant slip amount $v_0^{slip}=2V_wl_s/(H+2l_s)$ can be derived. 
Note this $v_0^{slip}$ tends to vanish as $H$ is sufficiently large.

\subsection{Power-law slip profile}\label{power}

\begin{figure}
\centerline{\psfig{figure=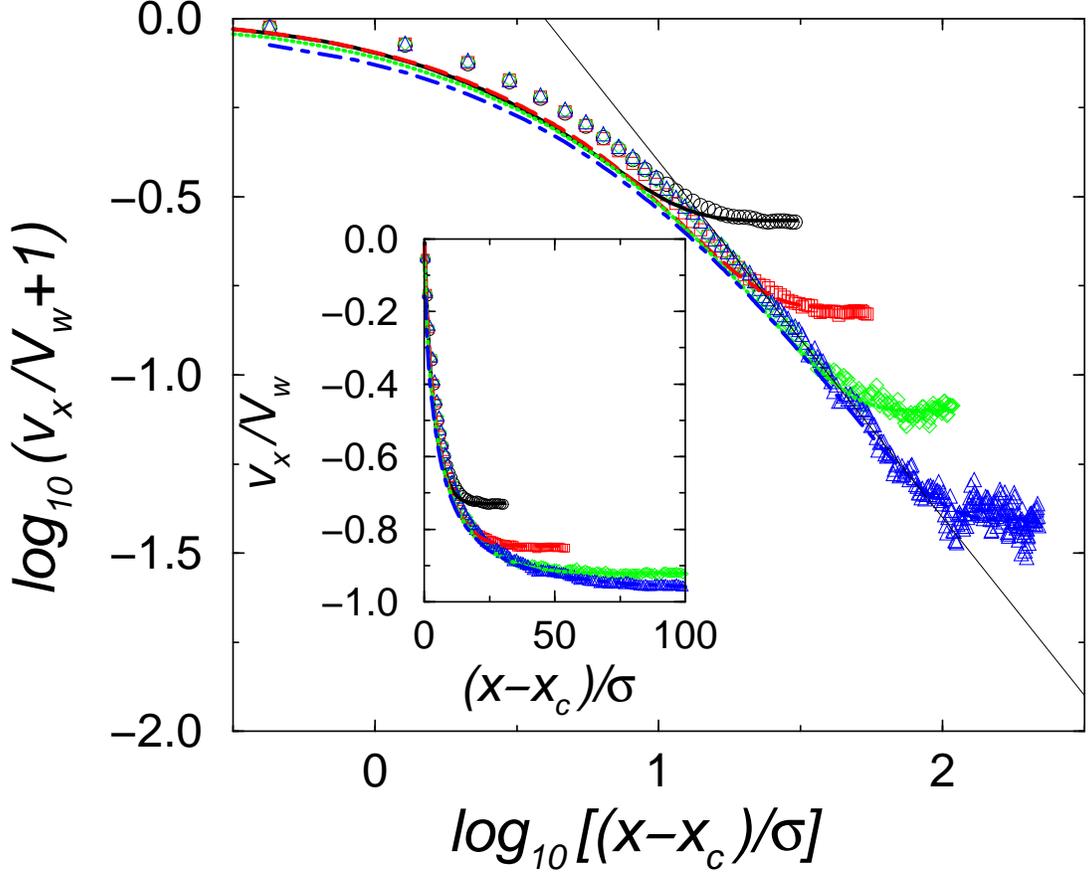,height=12.cm}}
\bigskip
\caption{Log-log plot of the slip profiles showing the power-law behavior. 
Here $v_x/V_w+1$ is the scaled slip velocity in the lower BL, 
and $(x-x_c)/\sigma$ measures the distance from the corner,
with $x_c=-L/2$ being the coordinate of the left corner and $x$ ranging
from $x_c$ to $0$ (the cavity center).
The lower wall is moving at $-V_w$, hence $v_x/V_w+1=1$
means complete slip and $v_x/V_w+1\rightarrow 0$ means no slip.
The MD results were obtained from four simulations for the same
$V_w=0.25\sqrt{\epsilon/m}$ but different $H$.
The symbols represent the MD results and the lines represent 
the continuum results, obtained for 
$H=13.6\sigma$ (circles and solid line), 
$H=27.2\sigma$ (squares and dashed line), 
$H=54.4\sigma$ (diamonds and dotted line), 
$H=108.8\sigma$ (triangles and dash-dotted line).
The thin straight line has the slope of $-1$, indicating that the $1/(x-x_c)$ 
behavior is approached for increasingly larger $H$. 
For $H=108.8\sigma$, the power-law behavior extends from 
$x-x_c\approx 14\sigma\approx 6l_s$ to $80\sigma\approx 35l_s$. 
Inset: The scaled tangential velocity $v_x/V_w$ in the lower BL,
plotted as a function of $(x-x_c)/\sigma$.
}\label{fig-power}
\end{figure}

Now we turn to the variation of the slip velocity along 
the solid surface, still in the regime of small Reynolds number 
(the largest ${\cal R}\approx 13$ for the largest $H$).
We have performed a series of MD simulations using the same wall speed
$V_w=0.25\sqrt{\epsilon/m}$ but different system size ($L$ and $H$).
The tangential slip velocity profiles at the fluid-solid interface, 
i.e., the slip profiles, are shown in the inset to Fig. \ref{fig-power}.  
Regardless of the distance $H$, there is always a small core region
in the immediate vicinity of the corner, on the order of a few $l_s$, 
where the slip amount displays sharp decay.
In particular, the structure of this core region is nearly independent
of the system size. As $H$ increases, however, 
a much slower variation of the slip profiles becomes apparent away from
the core region. To find out the nature of this slow variation, 
we plot in Fig. \ref{fig-power} the same data in the log-log scale. 
The dashed line has the slope of $-1$, indicating a power-law behavior:
away from the core region there is a wide partial-slip region in which
the amount of slip varies as $1/(x-x_c)$ 
where $x_c$ is the $x$ coordinate of the corner.
According to the NBC, the power-law variation of slippage means
the same variation of tangential stress.
Therefore, the asymptotic $1/r$ behavior of the stress variation 
\cite{batchelor,hua} has indeed been observed in MD simulations. 
With $H$ being finite and $L$ being sufficiently large ($L\ge 4H$), 
far from the corners there is always a region of uniform shear flow, where 
the slip amount is a constant, given by $v_0^{slip}=2V_wl_s/(H+2l_s)$. 
This is seen from the inset to Fig. \ref{fig-power} where
each slip profile shows a plateau.   
The small constant $v_0^{slip}$ (for $H\gg l_s$) acts as an outer cutoff on 
the $1/(x-x_c)$ profile at $x-x_c\sim H$.
For our largest MD simulation with $H=108.8\sigma$ (${\cal R}\approx 13$), 
the $v^{slip}\propto 1/(x-x_c)$ behavior actually extends to 
$\approx 80\sigma$ (or $\approx 35l_s$).  
Therefore, as $H\rightarrow\infty$ and $v_0^{slip}$ approaches $0$ 
(no-slip), the power-law region can extend to hundreds of $l_s$
or even more. A large power-law partial-slip region is significant, 
because the outer cutoff length scale directly determines 
the integrated effects, such as the total steady-state dissipation.  
While in the past the asymptotic $1/r$ stress variation away from the corner 
or the moving contact line has been known in continuum hydrodynamics
\cite{batchelor,hua}, to our knowledge the observation 
that the partial slip is of the same spatial dependence has not been 
previously reported.

The continuum results are also shown in Fig. \ref{fig-power}
for comparison. They were calculated using the same set of 
material parameters $\rho$, $\eta$, and $l_s$
corresponding to the same local properties in all the four MD simulations.
The overall agreement with the MD results is excellent. This not only
clearly demonstrates the validity of our continuum model, but also
confirms the power-law partial-slip region in the continuum hydrodynamics.

\subsection{Large Reynolds number}\label{large-reynolds}

\begin{figure}
\centerline{\psfig{figure=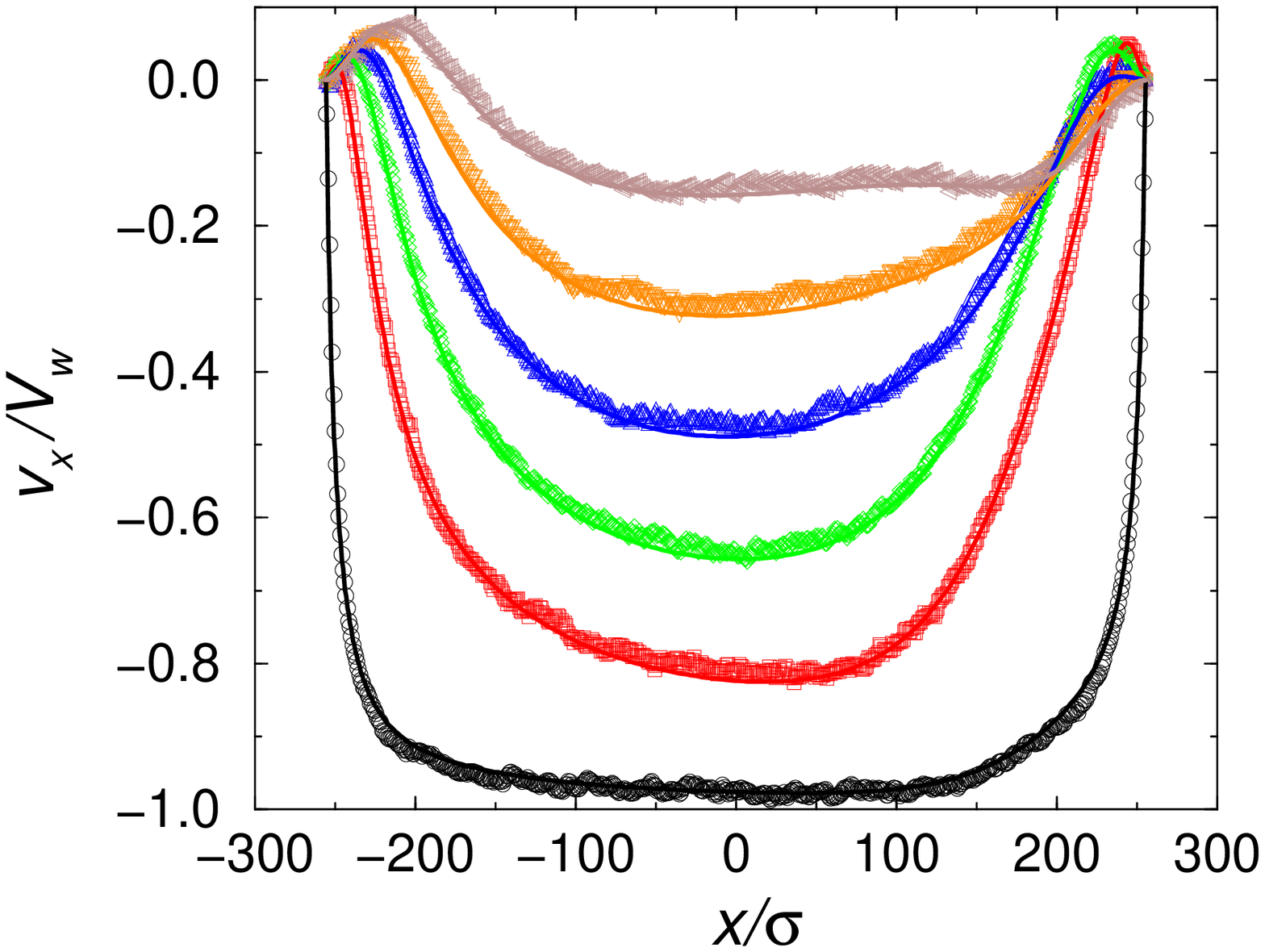,height=12.cm}}
\bigskip
\caption{Comparison between the MD and continuum 
results for a flow at ${\cal R}\approx 50$. 
The symbols denote the MD profiles of $v_x/V_w$ at different $z$ levels,
obtained for $V_w=0.5\sqrt{\epsilon /m}$, $L=511\sigma$, and $H=204\sigma$. 
The $v_x/V_w$ profiles are symmetric about the center plane $z=H/2$,
hence only the lower half is shown for $z=0.425\sigma$ (circles), 
$17.425\sigma$ (squares), $34.425\sigma$ (diamonds), 
$51.425\sigma$ (up triangles), $68.425\sigma$ (down triangles), 
and $85.425\sigma$ (left triangles).
The solid lines denote the corresponding continuum profiles.
}\label{fig-reynolds}
\end{figure}

MD simulations have been carried out to investigate the flow fields at 
large Reynolds number. Figure \ref{fig-reynolds} shows
the MD profiles of $v_x/V_w$ at different $z$ levels, obtained from
a large-scale simulation for $V_w=0.5\sqrt{\epsilon /m}$, $L=511\sigma$, 
and $H=204\sigma$ (${\cal R}\approx 50$). The approximate
fore-aft symmetry of the Stokes flow disappears.
In particular, the slip profiles associated with the left and right
corners are no longer symmetric, especially in the partial-slip regions
where the $1/r$ variation would appear were the Reynolds number being small.
This deviation from the power-law behavior described in Sec. \ref{power} 
is due to the vorticity convected with the fluid.
The continuum profiles of $v_x/V_w$ are also shown in Fig. 
\ref{fig-reynolds}. Excellent agreement is seen,
from the fine features in the molecular-scale vicinity of the solid walls
to the global flow.

\section{Concluding remarks}

In summary, we have carried out MD simulations to study the fine structure 
of the driven cavity flow. It has been verified that the Navier  
boundary condition can quantitatively describe the fluid slipping
at the solid wall. It has also been shown that   
close to the corner, where velocity variation is extremely fast, 
the shear stress is still Newtonian. Based on these MD facts, 
a continuum hydrodynamic model has been formulated. This model involves 
no adjustable parameter, and it can produce flow fields 
in quantitative agreement with those from MD simulations.

Recently, people have developed some MD-continuum hybrid methods for 
the study of fluid dynamics that involves complex small-scale structure
where validity of continuum formulations is not clear
\cite{thompson,hadji,chensy1,chensy2}. These hybrid methods have been
successfully applied to study the moving contact line \cite{hadji}, 
channel flow with nano-scale rough wall \cite{chensy1}, 
and the corner singularity in driven cavity flow \cite{chensy2}. 
Here we want to point out that when formulated correctly,
continuum hydrodynamic approach may still be applicable to some of these
problems, including the moving contact line \cite{qws} and 
the driven cavity flow. For each problem, the validity of 
its continuum model has been verified, first by a direct MD measurement 
and then by a comparison with full MD results.

\section*{Acknowledgment}
This work was partially supported by RGC DAG 03/04.SC21.

\begin{figure}
\centerline{\psfig{figure=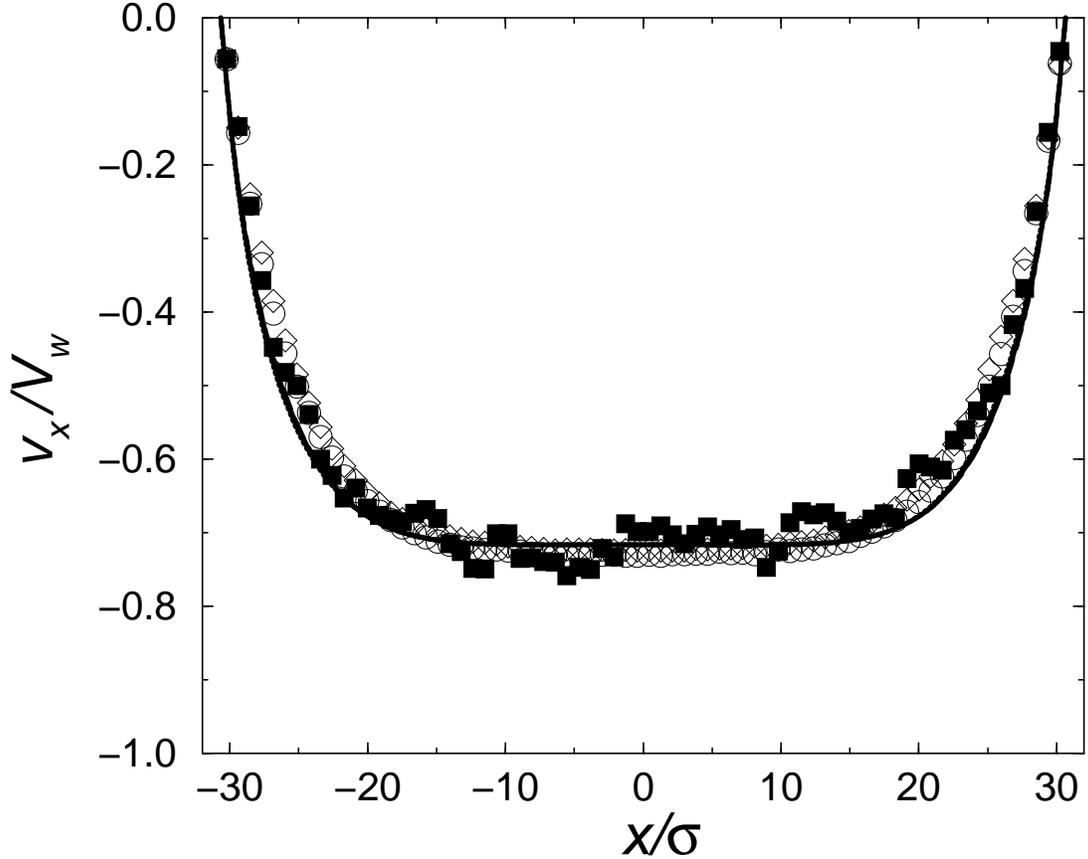,height=12.0cm}}
\bigskip
\caption{Boundary-layer tangential velocity profiles.
The scaled tangential velocity $v_x/V_w$ in the lower BL
is plotted as a function of $x/\sigma$. 
(That for the upper BL can be obtained by symmetry.) 
The lower wall is moving at $-V_w$, hence $v_x/V_w=0$
means complete slip and $v_x/V_w=-1$ means no slip.
Close to the corner (within a few $\sigma$) 
the slip is near-complete while away from the corner the slip becomes smaller.
There is a uniform shear flow in the central region where 
the slip is a constant.  
The three cases shown here are of the same $L=61.3\sigma$ and
$H=13.6\sigma$ but different $V_w$. 
The circles, squares, and diamonds denote 
the MD results for $V_w=0.25\sqrt{\epsilon /m}$, 
$0.01\sqrt{\epsilon /m}$, and $1.25\sqrt{\epsilon /m}$,
respectively. (The squares for the smallest $V_w$ 
show the largest statistical fluctuation.) It is seen that
the three MD profiles show negligible deviation from each other.
The solid, dashed, and dotted lines 
denote the continuum results 
for $V_w=0.25\sqrt{\epsilon /m}$, 
$0.01\sqrt{\epsilon /m}$, and $1.25\sqrt{\epsilon /m}$,
respectively. They also show negligible deviation from 
each other.
}\label{fig-slip}
\end{figure}

\begin{figure}
\centerline{\psfig{figure=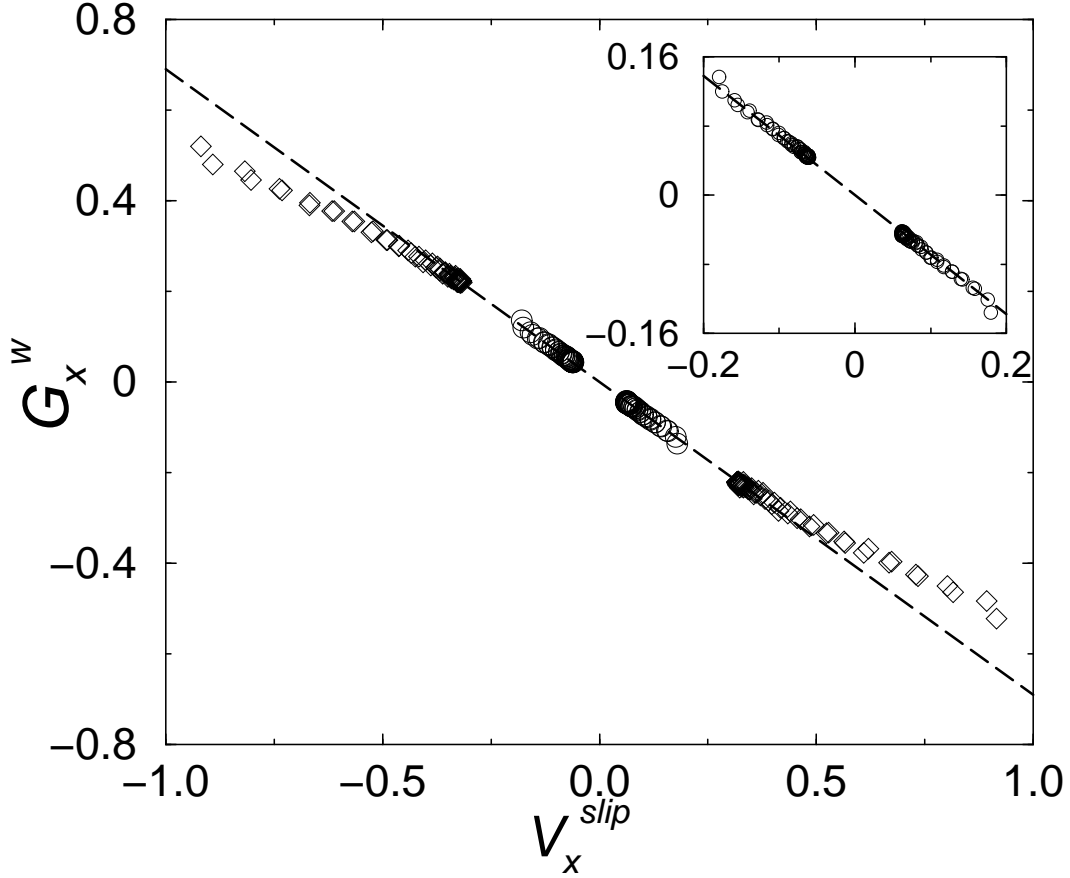,height=12.0cm}}
\bigskip
\caption{The tangential wall force $G_x^w$ 
(in the unit of $\epsilon/\sigma^3$) is plotted as a function
of the slip velocity $v_x^{slip}$ 
(in the unit of $\sqrt{\epsilon /m}$) for the lower 
($v_x^{slip}>0$) and upper ($v_x^{slip}<0$) BL's.
The two cases shown here are of the same $L=61.3\sigma$ and
$H=13.6\sigma$ but different $V_w$. The circles and   
diamonds denote the cases of $V_w=0.25\sqrt{\epsilon /m}$
and $1.25\sqrt{\epsilon /m}$, respectively. 
The dashed line is for eye guidance.
Linearity is seen to be well preserved for $V_w=0.25\sqrt{\epsilon /m}$
(circles, see also the inset for an enlarged plot). The slip coefficient
$\beta$ is obtained to be $0.69\sqrt{\epsilon m}/\sigma^3$.
For $V_w=1.25\sqrt{\epsilon /m}$ (diamonds),
however, nonlinearity shows up for $|v_x^{slip}|>0.5
\sqrt{\epsilon /m}$.
}\label{fig-wall}
\end{figure}

\begin{references}
\bibitem{note-nonlinear}
For $V_w=1.25\sqrt{\epsilon /m}$, the slip coefficient $\beta$ (or
equivalently the slip length $l_s=\eta/\beta$) is no longer a constant  
when $|v_x^{slip}|>0.5\sqrt{\epsilon /m}$ (see Fig. \ref{fig-wall}). 
However, this nonlinearity does not show up appreciably in the slip profile.
As seen in Fig. \ref{fig-slip}, the MD profile of $v_x/V_w$ for 
$V_w=1.25\sqrt{\epsilon /m}$ shows negligible deviation from that
for $V_w=0.25\sqrt{\epsilon /m}$, and is well reproduced by the
continuum model with a constant slip length.
\end{references}
\end{document}